\documentclass[twocolumn]{aastex701}

\usepackage{graphicx}	% Including figure files
\usepackage{amsmath}	% Advanced maths commands
\usepackage{amssymb}	% Extra maths symbols

\usepackage{multirow}
\usepackage{subcaption}
\usepackage{xcolor}

\newcommand{\psrpoppy}{\textit{PsrPopPy}}
\newcommand{\startrack}{\textit{StarTrack}}

\begin{document}
% \label{firstpage}
% \pagerange{\pageref{firstpage}--\pageref{lastpage}}
% \maketitle

%%%%%%%%%%%%%%%%%%% TITLE PAGE %%%%%%%%%%%%%%%%%%%

\title{Empirical determination of the Galactic neutron star--black hole merger rate using StarTrack models}

% The list of authors, and the short list which is used in the headers.
% If you need two or more lines of authors, add an extra line using \newauthor
% \author[Pierre Jacques et al.]{
% T. Pierre Jacques$^{1,2,3}$\thanks{E-mail: tp0052@mix.wvu.edu},
% D.R. Lorimer$^{1,2}$,
% N. Pol$^{4,5}$,
% M.A. McLaughlin$^{1,2}$ \newauthor
% and K. Belczynski$^{6}$
% \\
% % List of institutions
% $^{1}$Department of Physics and Astronomy, West Virginia University, Morgantown, WV 26506-6315\\
% $^{2}$Center for Gravitational Waves and Cosmology, West Virginia University, Chestnut Ridge Research Building, Morgantown, WV 26505\\
% $^{3}$Department of Physics, University of Idaho, Moscow, ID 83843, USA\\
% $^{4}$Department of Physics, Oregon State University, Corvallis, OR 97331, USA\\
% $^{5}$Center for Gravitation, Cosmology and Astrophysics, Department of Physics, University of Wisconsin-Milwaukee,  Milwaukee, WI 53201, USA\\
% $^{6}$Deceased\\
% }

\author{Terrence Pierre Jacques}
\affiliation{Department of Physics and Astronomy, West Virginia University, Morgantown, WV 26506-6315}
\affiliation{Center for Gravitational Waves and Cosmology, West Virginia University, Chestnut Ridge Research Building, Morgantown, WV 26505}
\affiliation{Department of Physics, University of Idaho, Moscow, ID 83843, USA}
\email{tp0052@mix.wvu.edu}

\author{D. R. Lorimer}
\affiliation{Department of Physics and Astronomy, West Virginia University, Morgantown, WV 26506-6315}
\affiliation{Center for Gravitational Waves and Cosmology, West Virginia University, Chestnut Ridge Research Building, Morgantown, WV 26505}
\email{duncan.lorimer@mail.wvu.edu}

\author{Nihan Pol}
\affiliation{Department of Physics and Astronomy, Texas Tech University, Lubbock, TX 79409}
\affiliation{Department of Physics, Oregon State University, Corvallis, OR 97331}
\affiliation{Center for Gravitation, Cosmology and Astrophysics, Department of Physics, University of Wisconsin-Milwaukee, Milwaukee, WI 53201, USA}
\email{npol@ttu.edu}

\author{M. A. McLaughlin}
\affiliation{Department of Physics and Astronomy, West Virginia University, Morgantown, WV 26506-6315}
\affiliation{Center for Gravitational Waves and Cosmology, West Virginia University, Chestnut Ridge Research Building, Morgantown, WV 26505}
\email{maura.mclaughlin@mail.wvu.edu}

\author{K. Belczynski}
\affiliation{Deceased}
\email{}

\begin{abstract}
The discovery of a pulsar in a binary system with a black hole would provide a unique laboratory for testing general relativity in the strong-field regime and offer vital constraints on massive star evolution. We assess prospects for the detectability of such systems by utilizing the radio pulsar population and survey models within the modeling package \textit{PsrPopPy} and, as a proof of principle, two evolutionary models with different treatments of common-envelope phases from the stellar population synthesis code \textit{StarTrack}. For these two models, assuming the radio pulsar beaming model proposed by Tauris and Manchester and accounting for the effects of orbital motion on pulsar detectability, we calculate upper limits on the number of neutron star--black hole systems in the Galaxy and the corresponding merger rate given the lack of any radio-detected pulsar--black hole binary systems in current radio surveys. For the most constraining model, we find a Galactic merger rate of $\mathcal{R} < 4.3~\text{Myr}^{-1}$ (at 95\% confidence). These results are extrapolated to the expected horizon distance for the fifth LIGO observing run, leading to LIGO detection rate estimates that are consistent with the latest values reported by the LIGO-Virgo-KAGRA collaboration. Our results indicate that the increased sensitivity of FAST, MeerKAT, and DSA offers the best prospects yet for uncovering this elusive population.
\end{abstract}

% \begin{keywords}
% stars: neutron -- stars: black holes -- methods: statistical -- surveys: pulsars -- gravitational waves
% \end{keywords}
% % ABOVE NEEDS TO FIXED, TO NEW APJ FORMAT:
% \keywords{\uat{Galaxies}{573} --- \uat{Cosmology}{343} --- \uat{High Energy astrophysics}{739} --- \uat{Interstellar medium}{847} --- \uat{Stellar astronomy}{1583} --- \uat{Solar physics}{1476}}

%%%%%%%%%%%%%%%%%%%%%%%%%%%%%%%%%%%%%%%%%%%%%%%%%%

\section{Introduction}

In recent years the LIGO/VIRGO collaboration has observed gravitational wave emission from the merger of stellar mass binary black holes \citep{Abbott_2016, Abbott_2019}, binary neutron stars \citep{Abbott_2017, Abbott_2020}, and neutron star--black hole (NSBH) systems \citep{Abbott_2021}. These detections have enabled direct merger rate calculations for these systems \citep{Abbott_2023}, finding the {local} NSBH merger rate ({redshift $z=0$}) to be in the range {9.1--84}~$\mathrm{Gpc}^{-3}~\mathrm{yr}^{-1}$ {\citep[][]{LIGO_GWTC_4}}. This result can be combined with binary population synthesis simulations to place constraints on the formation channels for these binaries \citep[see][for a review]{Mandel_2022LRR}. However, the precision of the NSBH merger rate implies a wide range in the prediction for the mass distribution function and formation channels for these systems \citep[e.g.,][]{Shao_2021, Broekgaarden_2021, Broekgaarden_2022, Roman_2021, Boco_2019, Kruckow_2018}.

{One may also test models of binary formation channels by placing limits on the number of NSBH systems in the Milky Way, and reconcile these limits with the lack of systems discovered in the Galaxy through radio pulsar surveys.}
While current and recent pulsar surveys of the Galaxy \citep[for a review, see][]{2022ASSL..465....1B} and globular clusters \citep[e.g.,][]{2021ApJ...915L..28P,2022A&A...664A..54G} have resulted in a significant increase in the pulsar sample, so far only one candidate NSBH binary has been found \citep[PSR~J0514--4002E in the globular cluster NGC~1851;][]{2024Sci...383..275B}. \citet{2023ApJ...951...20J} investigated the implications of the lack of confirmed systems on our understanding of the population of long-period ($>100$--1000~yr) NSBH systems, which may be present in the sample of long-period pulsars. They conclude that such systems cannot be currently ruled out, and that detections might be made in the next decade as timing observations improve.  

{
While NSBH systems in globular clusters will yield significant insight into the spacetime near stellar-mass black holes, these systems are likely to form through dynamical interactions within the dense cluster environment. As a result they are not expected to shed light on the formation channels for these systems through common-envelope evolution.
Since no pulsar companion within a NSBH system has been discovered in the Galactic field, most studies of the formation channels for compact binaries have focused on pulsar binaries discovered in pulsar surveys, with either a white dwarf \citep[e.g.,][]{nswd_popsynth_1,nswd_popsynth_2} or neutron star \citep{beniamini_popsynth_nopsrpop,vigna_gomez_popsynth_nopsrpop,shaughnessy_popsynth_psrpop} as the companion. As reviewed by \citet{2023pbse.book.....T}, these analyses usually evolve the binary system starting from zero-age main sequence stars, and track the evolution of the system as both stars undergo supernova explosions and eventually form the compact object binary system. Once the binary is formed, some analyses either compare the systems directly to those detected in pulsar surveys, ignoring the radio evolution of the pulsar emission in the compact binary \citep[e.g.,][]{beniamini_popsynth_nopsrpop,andrews_popsynth_nopsrpop,vigna_gomez_popsynth_nopsrpop}, or attempt to model the radio emission in a fairly simplistic manner \citep[e.g.,][]{oslowski_popsynth_psrpop,shaughnessy_popsynth_psrpop, Kinugawa_2022}. 
{While the formation of pulsars within NSBH binaries may prove to be dynamically challenging \citep{Liotine_2025}, some exceptions which model the radio detection of pulsars in these binaries include the studies presented in \citet{debatri_popsynth_fullpsrpop_2020} and \citet{debatri_popsynth_fullpsrpop_2021}, which model the spin evolution of pulsars within binaries. Similar to our preliminary study here, \citet{debatri_popsynth_fullpsrpop_2021} modeled the radio observation of pulsars within pulsar-black hole binaries, using the \textit{COMPAS} \citep{Stevenson_2017, compas2021, compas2025} population synthesis code, coupled to the \textit{PSREvolve} pulsar evolution code \citep{PSREvolve_2011, debatri_popsynth_fullpsrpop_2020}. \citet{Shao_2018} instead used the \textit{BSE} population synthesis code \citep{Hurley_2002} to also model radio observation of Galactic of pulsar-black hole systems, modeling the pulsar radio luminosity using the methods described in \citet{2006ApJ...643..332F}. The work presented in \citet{Sgalletta_2023} used the \textit{SEVN} population synthesis code \citep{SEVN_2017, SEVN_2019, SEVN_2020, SEVN_2023} to simulate observations of pulsars in binary neutron star systems, but coupled pulsar evolution and selection effects using \psrpoppy{}, a software suite written to model radio pulsar evolution, simulate observations of the Galactic distribution of pulsars.  In this preliminary study we a novel framework to combine the results from binary population synthesis codes with \psrpoppy{}, and accurately model the orbital degradation effects for binary pulsar systems \citep{Pol_2021}, which have not yet been included in previous studies.}
} 

{
We showcase this framework by connecting results from \startrack{} \citep[][]{BelczynskiA,BelczynskiB, Belczynski_2020} simulations of NSBH systems to \psrpoppy{}. 
\startrack{} allows us to simulate a population of NSBH systems, starting from the zero-age main sequence mass and metallicity for each star in the binary system, through the binary evolution, including various prescriptions for mass transfer and common-envelope (CE) phase evolution, to the present-day orbital parameters for the NSBH system. These orbital parameters for the NSBH system can then be passed to \psrpoppy{}, where they can be coupled with known pulsar population properties and radio telescope responses to determine if the system will be detectable in a given radio pulsar survey. The results from these simulations can then be directly compared with the samples found in real pulsar surveys, allowing us to probe how different choices in the evolution of the binary system affect the overall yield of these systems in pulsar surveys. Furthermore, we  translate the expected yield of NSBH systems in our Galaxy to the corresponding Galactic NSBH merger rate. A comparison of this merger rate with that measured by LIGO allows us to further constrain NSBH evolutionary channels. }

The rest of this paper is organized as follows. In Section 2 we explain how we connect \startrack{} with \psrpoppy{} and build the statistical framework for determining the probability distributions of the Galactic merger rate and the number of NSBH systems. We present the results of this study in Section 3 and discuss and place them in context with other work in Section 4.

\section{The population of pulsar--black hole binary systems}

\begin{figure*}
\includegraphics[width=0.9\textwidth]{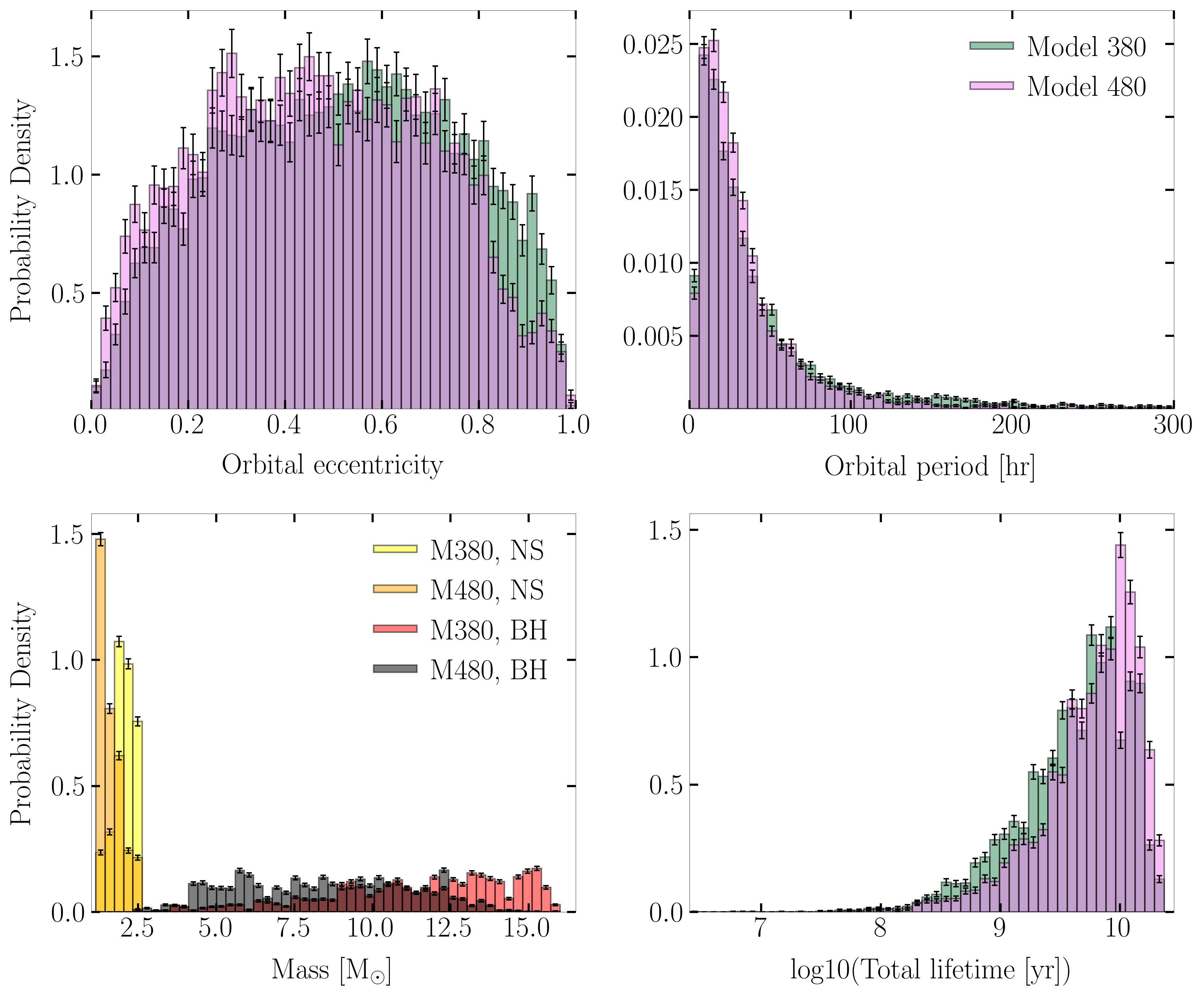}
\caption{Distributions of NSBH systems within models 380 and 480 from \startrack{}: orbital eccentricities (top left), orbital periods (top right), masses (bottom left) and time prior to coalescence due to gravitational wave energy loss (bottom right). }
\label{fig:physical}
\end{figure*}

\begin{figure}
\includegraphics[width=8cm]{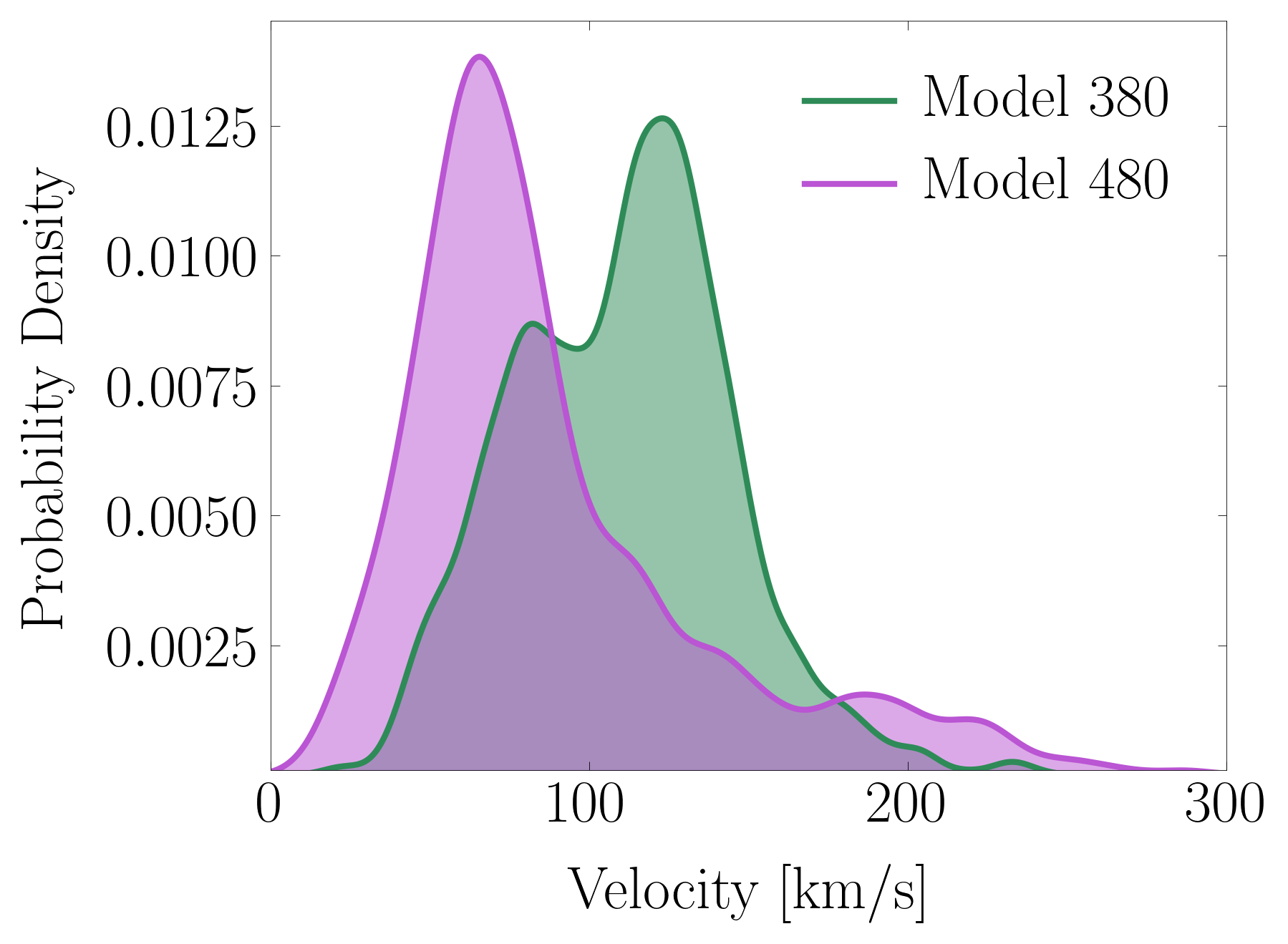}
\caption{Kernel density estimations of the velocity distributions from the \startrack{} models. Model 380 has a mean of 111~km~s$^{-1}$, while model 480 has a mean of 89.5~km~s$^{-1}$.}
\label{fig:3}
\end{figure}

Our overall approach is to create a synthetic population of NSBH systems, and use their resulting parameter distributions to simulate the detections of these systems by radio pulsar surveys. By characterizing the detection rate of these surveys, as described below, we can apply a Poissonian statistical framework to constrain the putative population and merger rate of NSBH systems.

\subsection{Creating a NSBH binary population}

To obtain realistic characteristics of a NSBH binary system population within the Galaxy, we make use of the population synthesis models in \startrack{} \citep{BelczynskiA,BelczynskiB, Belczynski_2020}. \startrack{} has undergone a number of improvements over the years, summarized in \citet{Olejak_2020}. Here we use the latest version presented in \citet{Olejak_2020}, with the additional modifications presented in \citet{Olejak_2021}. Specifically, we adopt two of the models used in \citet{Olejak_2021}, M380.B and M480.B, where B denotes the submodel. {Given that this paper serves as initial test bed for our novel analysis techniques, we limit this study to just these models. Future applications of our methodology should include a larger number of population synthesis models, to ultimately place constraints on the dominant physical processes leading to the formation of NSBH systems.} Both models use the delayed supernovae engine mechanism described in \citet{Fryer_2012, Belczynski_2012}. In submodel B donor stars within the Hertzsprung gap merge with their companion during the CE phase. For simplicity, we drop the M and B labels in the remainder of this paper. Model 380 includes standard treatment of Roche lobe overflow (RLOF), consistent with previous \startrack{} models, while  model 480 includes the updated RLOF treatment presented in \citet{Olejak_2021}. Specifically, it incorporates a revised CE criterion that includes additional conditions to apply the instabilities described in \citet{Pavlovskii_2017}. For model 480, CE development may only occur once the conditions outlined in \S3.1 of \citet{Olejak_2021} are satisfied, modeling one of the two instabilities for RLOF, otherwise the CE does not develop and the code models stable RLOF. These requirements effectively add more restrictions to the onset of the CE phase. We show the cumulative effects of the differences in these models on a synthetic population of NSBH systems in Fig.~\ref{fig:physical}, plotting histograms of orbital eccentricity, orbital period, neutron star {and black hole} mass and total system lifetime. We observe that the greatest difference between the two models is in the distribution of neutron star {and black hole} mass and system lifetime. In Fig.~\ref{fig:3} we show the kernel density estimations of the final system velocity, after the second supernova, from each system. There we observe a distinct difference in the mean velocity between the two models, with model 380 having a mean of 111~km~s$^{-1}$, while that of model 480 is 89.5~km~s$^{-1}$.

\begin{table*}
\centering
\caption{Summary of {\it PsrPopPy} model specifications for the pulsar survey parameters considered in this work. From left to right, the surveys considered are: Parkes \citep{2001MNRAS.328...17M}, High Time Resolution Universe (HTRU) low- and mid-latitude surveys \citep[][respectively]{2010MNRAS.409..619K,2019MNRAS.484.5791B}, Pulsar Arecibo L-Band Feed Array \citep[PALFA,][]{2006ApJ...637..446C} and the Green Bank Northern Celestial Cap survey \citep[GBNCC,][]{2014ApJ...791...67S}. We also list model specifications for ongoing and future surveys conducted with DSA \citep{2019BAAS...51g.255H}, FAST \citep{2021RAA....21..107H}, and MeerKAT \citep{2024MNRAS.531.3579T}.}
\begin{tabular}{cccccccccc}
\hline \hline
    & &\multicolumn{8}{c}{Pulsar Survey} \\
Parameter & Unit & Parkes & HTRU low & HTRU mid & PALFA & GBNCC & DSA & FAST & MeerKAT \\ \hline
{ Digitization} factor & & 1.2 & 1.2 & 1.2 & 1.1 & 1.3 & 1.0 & 1.0 & 2.0\\ 
Antenna gain & K/Jy & 0.6 & 0.6 & 0.6 & 8.5 & 2.0 & 10& 16 & 2.8 \\ 
Integration time & s & 2100 & 340 & 540 & 268 & 120 & 900& 300 & 600 \\ 
Sampling time & $\mu$s & 250 & 64 & 64 & 64 & 82 & 100& 49 & 74 \\ 
System temperature & K & 25 & 25 & 25 & 25 & 46 & 25 & 25 & 18 \\ 
Centre frequency & MHz & 1374 & 1352 & 1350& 1350& 350 & 1300 & 1250 & 1284 \\ 
Bandwidth & MHz & 288 & 340 & 340 & 340& 100 & 1300 & 1300& 776 \\ 
Channel bandwidth & kHz & 3000 & 390 & 300& 300 & 24.4 & 134& 244 & 757 \\ 
Beam width & arcmin & 14 & 14 & 14 & 3.6 & 30 & 0.06& 3 & 1.7\\ 
Min declination & deg & --90 & --90 & --90 & 0 & 38 & --37& --14 & --90 \\ 
Max declination & deg & 27& 90 & 27& 38 & 90 & 90 & 65 & 40 \\ 
Min Galactic longitude & deg & --100 & --80 & --120 & 32 & --180 & --180 & --180 & --180 \\ 
Max Galactic longitude & deg & 50 & 30 & 30 & 77 & 180 & 180 & 180 & 10\\ 
Max $|$Galactic latitude$|$ & deg & 5 & 3.5 & 15 & 5 & 90 & 90 & 10 & 5\\ 
\hline
\end{tabular}
\label{tab:1}
\end{table*}

\subsection{Simulating pulsar evolution and detection}
\label{pulsar_evol_detect}

To simulate radio observations of pulsars in these systems, we generate pulsar populations using the freely available  
\textit{PsrPopPy}\footnote{The version of the code we use can be found at https://github.com/devanshkv/PsrPopPy2} software \citep[][hereafter BLRS14]{PSRPopPy}. Specifically, we use the \textit{Evolve} and \textit{DoSurvey} programs to generate and detect model pulsars. 
%In the summary below, we refer the reader to specific
%sections of BLRS14 for further details. 
While we provide a brief summary below, we refer the reader to specific sections of BLRS14 for further details.
Much of the implementation in \textit{PsrPopPy} is motivated by the earlier
population study of \citet{2006ApJ...643..332F}.

We first use the \textit{Evolve} program to simulate the positions, spins and luminosities of each pulsar within a given population. We extract the ages, orbital parameters and final system velocity for each binary from \startrack{}, restricting our analyses to systems that merge within a Hubble time, where the inspiral phase is driven by gravitational wave emission. The model pulsars are given an initial position within the Galaxy before being subsequently evolved through the Galactic potential (\S 3.4 of BLRS14), while we set their birth velocity to be the final system velocity drawn from the \startrack{} data. To model pulsar spin evolution, we assume a magnetic dipole model as described in detail \S 3.3 of BLRS14. In brief, each pulsar is assigned an initial spin period and magnetic field using the parameters summarized in Table 2 of BLRS14. These are then evolved in time using Equation 17 from BLRS 14 until the neutron star age, randomly drawn from a \startrack{} population model, is reached. As a result of this process, each model pulsar has a present-day spin period, $P$, and time derivative, $\dot{P}$.
\textit{Evolve} also accounts for the radio beaming fraction $f_b$ directly, which is a simple function of $P$
\citep[Equation 15 of][]{1998MNRAS.298..625T} and we only consider pulsars that are beaming toward Earth. 
As a result, we do not involve further beaming corrections in our rate calculations below. We also
only consider those pulsars that are deemed to be radio loud and above the death line
(Equation 20 of BLRS14).
For those pulsars that are potentially observable, radio luminosities are then calculated using a power-law expression that is a function of $P$ and $\dot{P}$ (\S 2.3.3 of BLRS14).

After obtaining a population of time evolved pulsars, we apply \textit{PsrPopPy's} \textit{DoSurvey} program, simulating the following surveys with model specifications summarized in Table~\ref{tab:1}; Parkes \citep{2001MNRAS.328...17M}, High Time Resolution Universe (HTRU) low- and mid-latitude surveys \citep[][respectively]{2010MNRAS.409..619K,2019MNRAS.484.5791B}, Pulsar Arecibo L-Band Feed Array \citep[PALFA,][]{2006ApJ...637..446C} and the Green Bank Northern Celestial Cap survey \citep[GBNCC,][]{2014ApJ...791...67S}. These comprise the largest and most sensitive sky surveys carried out to date at radio frequencies. { The detection thresholds for these surveys were calculated using the radiometer equation as described in \S 4.1.2 of BLRS14. These calculations incorporate the effects of multi-path scattering in the interstellar medium, as well as instrumental broadening and finite-level quantization  in the digitized survey data.

In reality, due to significant orbital motion of the binary system during a survey observation, the signal-to-noise ratios from these calculations need to be scaled to account for the Doppler smearing that degrades the pulsar signal. We follow the procedure described by \citet{Pol_ucb} and, for each survey observation of a model { pulsar}, compute an orbital degradation factor. This is a multiplicative quantity between zero and one, where a lower degradation factor indicates more significant Doppler smearing during the observation. {Crucially, the \citet{Pol_ucb} framework explicitly accounts for acceleration searches by evaluating signal recovery for both unaccelerated and accelerated search routines, appropriately averaged over the full orbit to account for the time spent at a given orbital phase. Higher-order techniques, such as jerk searches \citep{2018ApJ...863L..13A}, are omitted from our sensitivity calculations because they have not been systematically applied across the historical large-scale radio surveys modeled in this work.} 
All signal-to-noise ratios computed using this framework are then multiplied by the appropriate degradation factor. By following this methodology we are able to use \startrack{} and \textit{PsrPopPy} in a compatible fashion; \startrack{} allows us to obtain a realistic population of NSBH systems, while \textit{PsrPopPy's} \textit{Evolve} program allows us to evolve their spin and magnetic fields, while traversing the Galactic potential, determining which types of these systems may contain observable pulsars.

\subsection{Probability distributions for $N_{\rm tot}$ and $\mathcal{R}$}
\label{n_tot_R}

We follow the analysis outlined in \citet{Kim_2003}, deriving probability distributions for the total number of NSBH systems in the Galaxy and the Galactic merger rate. 
We consider a synthetic population of NSBH binary systems totaling $N_{\rm tot}$, assuming the neutron stars may be observed as pulsars via radio surveys. We assume their detection as radio band pulsars to be a Poisson process. The probability of finding some observed number of pulsars $N_{\rm obs}$ in any given survey 
\begin{equation} \label{eq1}
    P \left(N_{\rm obs}; \lambda \right) = \frac{\lambda^{N_{\rm obs}} e^{-\lambda}} { N_{\rm obs}! },
\end{equation}
where  $\lambda$ is the expected number of observed pulsars, i.e. $\lambda \equiv \left \langle N_{\rm obs} \right \rangle$. 
{ We assume a uniform (uninformative) prior on $\lambda$.}
Given that there have not been any radio observations of Galactic pulsars with BH companions to date, we set $N_{\rm obs} = 0$, so that { the resulting posterior}
    $P \left(\lambda \right) = e^{-\lambda}$.
As discussed by \citet{Kim_2003}, varying the total number of systems generated in our synthetic population across several simulations leads to a linear correlation between $N_{\rm tot}$ and $\lambda$, so that
\begin{equation} \label{eq3}
    \lambda = \alpha N_{\rm tot},
\end{equation}
where $\alpha$ is a constant { reflecting the observational selection effects given} our choice of luminosity distribution, scale height, beaming correction factor and orbital parameters discussed in the previous section.
To derive a PDF describing the total number of NSBH systems in the Galaxy, we apply Bayes's theorem to find
\begin{equation} \label{eq5}
    P \left( N_{\rm tot} \right) = P \left( \lambda \right) \left| \frac{d\lambda}{d N_{\rm tot}} \right| = \alpha e^{-\alpha N_{\rm tot}}.
\end{equation}
Here we note that the probability distribution for $N_{\rm tot}$ is a function of $\alpha$. 
To obtain { the probability} distribution of NSBH merger rates{, $P(\mathcal{R})$, we perform Monte Carlo sampling from $P(N_{\rm tot})$ and $P(\tau)$ via}
\begin{equation} \label{eq8}
    \mathcal{R} = \frac{N_{\rm tot}}{\tau}, 
\end{equation}
where $\tau$ is { the total lifetime drawn from} the distribution shown in the lower-right panel of Fig.~\ref{fig:physical}. { This lifetime represents the duration} the pulsar { evolves} in the Galactic gravitational potential { from} birth { until} the system merges due to gravitational wave emission{, as} computed by \startrack{}.

\subsection{Calculating probability distributions}

To calculate $P(N_{\rm tot})$ and $P(\mathcal{R})$, we use $N_{\rm tot}$ as a free parameter to calculate $\alpha$, via Equation \ref{eq3}. 
%While we fix our distribution of neutron star lifetimes from our\startrack{} simulation, we vary $N_{\rm tot}$, randomly assigning each pulsar an age from the distribution. 
Given the distribution of neutron star lifetimes from the \startrack{} simulations shown in Fig.~\ref{fig:physical}, we assign each pulsar in $N_{\rm tot}$ an age that is drawn from this distribution.

\begin{figure}
\includegraphics[width=8cm]{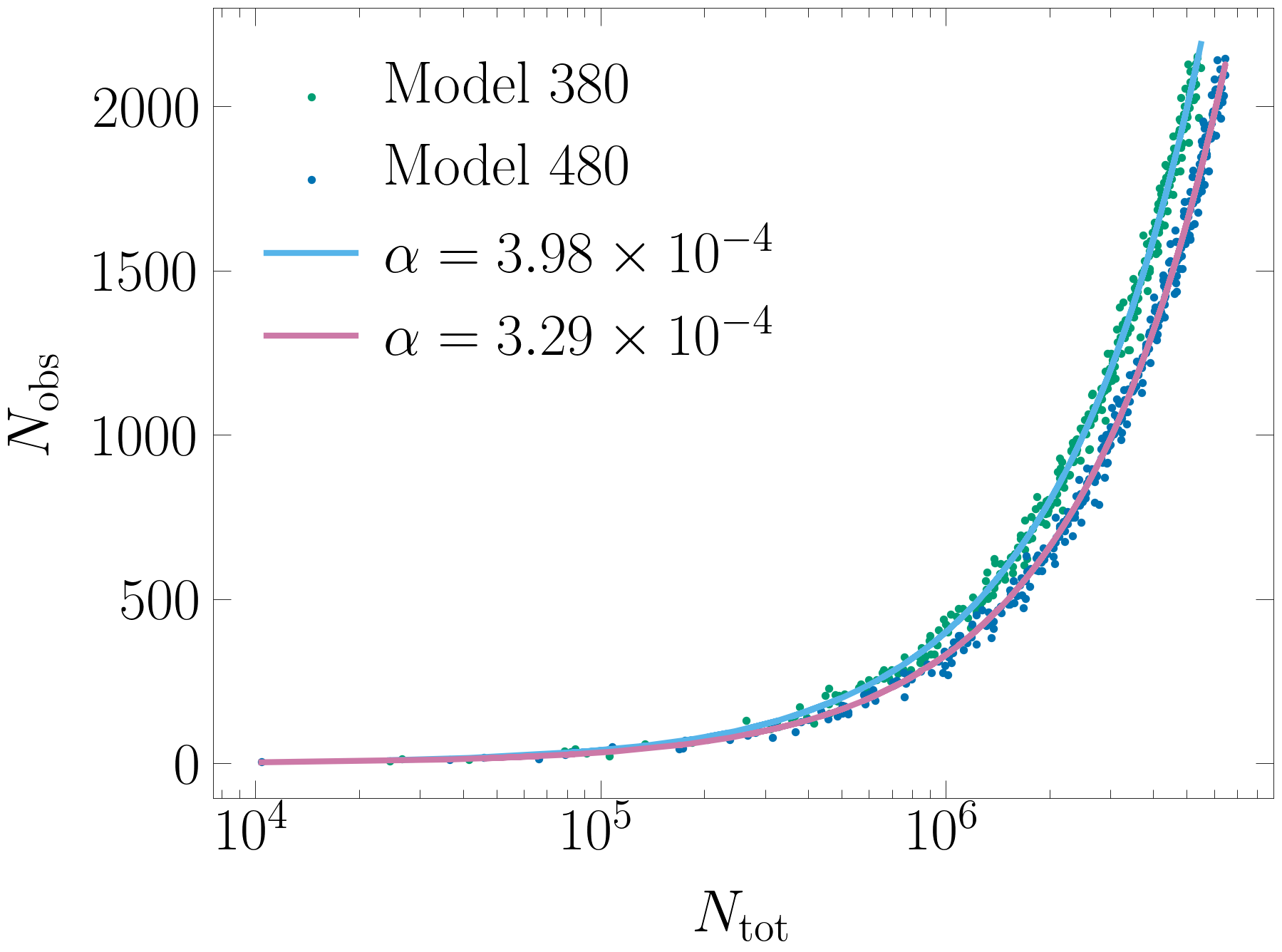}
\caption{Results of the \textit{DoSurvey} realizations of $N_{\rm tot}$, tracing out the linear correlation between $N_{\rm tot}$ and $N_{\rm obs}$. We use a standard least-squares fit to find the correlation parameter $\alpha$.}
\label{fig:2}
\end{figure}

We then apply the set of surveys given in Table~\ref{tab:1} to this population, and sum over the number of pulsars that are detected. As specified by Equation~\ref{eq3}, we find a linear relationship between $N_{\rm tot}$ and $N_{\rm obs}$ in Fig.~\ref{fig:2}. We note that our simulations require quite large numbers of pulsars to be generated for only a relatively few detections. This confirms that our analyses are accounting for the difficulty in detecting these systems, either from the fact that the pulsars are not beaming towards Earth, they are too dim to be observed, from Doppler smearing, ISM effects, or weak emission, or the orbital degradation may render these systems undetectable.

\section{Results}
\label{results_discussion}

\subsection{Total Galactic coalescence rate}

To calculate the Galactic merger rate, we use Equation~\ref{eq8}, convolving the probability distributions for $\tau^{-1}$, taken from the \textit{StarTrack} data, with $N_{\rm tot}$, derived from our simulations in Section~\ref{n_tot_R}. The distribution for $N_{\rm tot}$ yields an upper limit on the number of NSBH systems we expect to exist in the Galaxy, given in Table~\ref{tab:2}. Model 380 yields an upper limit of $9,300$, while  model 480 yields $11,200$, both at the 95\% confidence interval. We omit the beaming fraction from this direct calculation, as it is already accounted for in the \textit{Evolve} program. We perform the convolution by applying a Monte-Carlo (MC) simulation to sample the rate distribution. Because these are upper limits, we find that the resulting distributions of the merger rates for both binary evolution models is exponential, peaking at zero with means of $1.2~\mathrm{Myr}^{-1}$ and $0.6~\mathrm{Myr}^{-1}$, for models 380 and 480, respectively. Through this MC simulation we sample the resulting probability distribution with $~5\times10^6$ points. This results in acquiring upper limits on the Galactic merger rate of $4.3~\mathrm{Myr}^{-1}$ and $2.3~\mathrm{Myr}^{-1}$.

\subsection{Detection rate for LIGO}

To calculate the detection rate for LIGO, we apply a similar treatment as \citet{Pol}, extrapolating our Galactic merger rates from the previous section to the observing volume of O5. From Equation 15 of \citet{Pol}, we have 
\begin{equation}
\begin{split}
    \mathcal{R}_{\rm LIGO} = 7.4\times10^{-3} & \left( \frac{\mathcal{R}}{(10^{10} L_{B, \odot} )^{-1} {\rm Myr}^{-1} } \right) \\
    & \times \left( \frac{D_h}{100~\rm{Mpc}} \right)^3 {\rm yr}^{-1},
\end{split}
\end{equation}
where $\mathcal{R}$ is the Milky Way merger rate weighted by the Milky Way B-band luminosity, $D_h$ is the LIGO horizon distance, and $L_{B, \odot} = 2.16\times10^{33}$ ergs~s$^{-1}$. {We  follow the prescription provided by \citet{Pol_2021}, which uses the horizon distance instead of the LIGO range distance $D_r$. The two are related by $D_h = 2.26 \times D_r$.} Weighting by the B-band luminosity is done to accurately account for the distribution of stars in the Galaxy. Following \citet{LIGONSBHDistance} we set $D_r$ to be $1.6$ times the range distance for binary neutron star systems. In this work consider the upper limit of the BNS range distance for the planned O5 observing run, estimated at 325 Mpc. This provides estimates for LIGO detection rates, again at the 95\% confidence interval, of $30.5~\mathrm{yr}^{-1}$ and $16.3~\mathrm{yr}^{-1}$ for models 380 and 480.

\section{Discussion}

In Fig.~\ref{fig:4} we compare our results for the NSBH merger rate, plotting the PDFs of both models as well as their upper limits at the 95\% confidence level, along with estimates at the 90\% credible interval from previous studies using \textit{StarTrack} \citep{Belczynski_2020, Olejak_2021}, \citet{Pol_2021}, along with the estimates derived from merger observations from the LIGO-Virgo-KAGRA (LVK) collaborations from  GWTC-3 \citep{Abbott_2023}, {which estimated the rate to lie in the range 7.8--140~$\mathrm{Gpc}^{-3}~\mathrm{yr}^{-1}$} and GWTC-4 \citep{LIGO_GWTC_4}. {We find our  results are compatible with current observations.} {At face value, these preliminary results imply} that the observed population of NSBH mergers most likely favored formation via a stable Roche lobe process. {However, we caution that} this paper represents an initial exploratory study, {and} our results cannot rule out any degeneracy between the models when compared to the GWTC-3 or GWTC-4 rates, as doing so would most likely require far more population models. {More pressingly, we limited this exploratory study to only two \startrack{} models, and thus we cannot confidently ascribe our results to the different treatments for CE mass transfer alone. Indeed, future studies will have to vary population synthesis models across natal kicks, CE efficiency, and star-formation, among other parameters, to confidently identify which model characteristics affect NSBH merger most strongly.} We also observe that the results from this study are more constraining than the upper limits found in \citet{Pol_2021}. This is most likely due to our consideration of more realistic NSBH system parameters, { whereas \citet{Pol_2021} used uninformative uniform priors on all the orbital parameters for the NSBH systems}. This indicates that future studies may also benefit from doing so, perhaps by considering variations on model 480.

{We also comment that this exploratory study does not distinguish between the broad class of NSBH systems observed by the LVK collaborations and those similar to the source of GW event GW230529 \citep{LVK_2024}. This may be of interest given that GW230529 was the first event to feature a primary binary component in the lower mass gap, with mass $3.6^{+0.8}_{-1.2} M_\odot$. Given that dynamical mass measurements of Galactic X-ray binaries have implied a mass gap between the heaviest neutron stars and the lightest black holes, GW230529 may well represent a new class of binary systems with a formation channel distinct from the larger population of NSBH binaries. However, we leave it to future studies to determine the impact such a sub-population of binaries may affect the derived merger rates using the methodology presented here. We note that plotting the NS and BH masses for each binary from \startrack{} in the lower left panel of Fig.~\ref{fig:physical}, we observe that both models predict very few objects in the mass gap. Thus, future studies can in principle explore how different CE prescriptions will affect this subpopulation of objects \citep[e.g.,][]{Shao_2021, Siegel_2023, Xing_2025}, and how their existence can affect Galactic BNS, BBH, and NSBH merger rates.}

\begin{table}
\centering
\caption{Upper limits on the number of NSBH systems in the Galaxy, as well as on the merger rate and the corresponding LIGO detection rate. These are given at the 95\% confidence level. From our study we also place upper limits on the expected number of pulsars in these systems to be observed by future surveys.}
\begin{tabular}{ccc}
\hline \hline 
{Parameter} & \multicolumn{2}{c}{ Model} \\
                   & 380 & 480 \\ \hline
Number of Galactic systems & 9251 & 11168 \\ 
Galactic merger rate ($\mathrm{Myr}^{-1}$) & 4.3 & 2.3 \\
LIGO detection rate ($\mathrm{yr}^{-1}$) & 30.5 & 16.3 \\ 
\hline 
{Future surveys} \\ \hline
DSA & 18 & 16 \\
FAST & 12 & 9 \\
MEERKAT & 1 & 0 \\ \hline
\end{tabular}
\label{tab:2}
\end{table}

\begin{figure*}%[t]
  \centering
  \includegraphics[width=\textwidth]{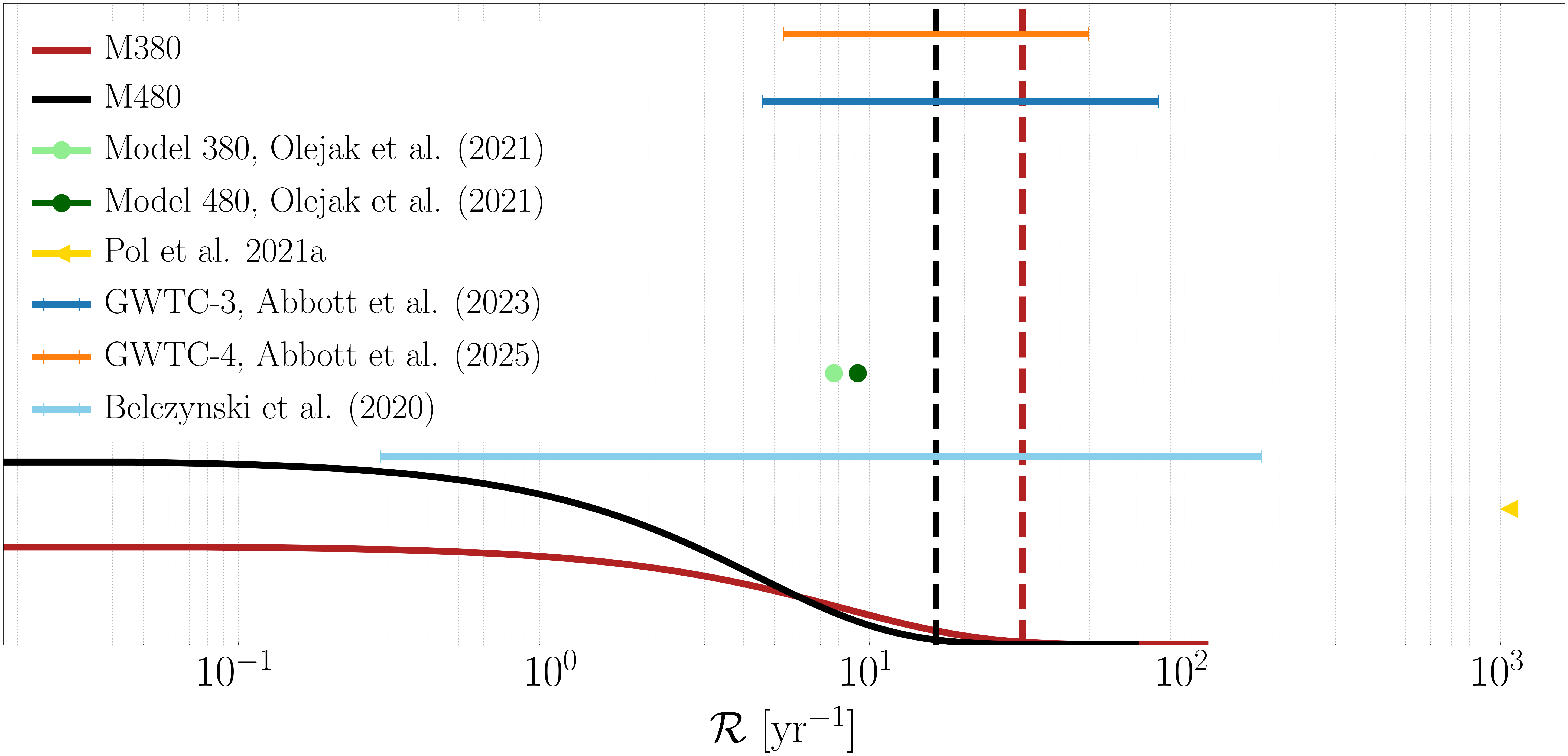}
\caption{Comparison of rates found in this work, \citet{Pol_2021} with those from other studies. { The dashed vertical lines mark the 95\% confidence level of our upper limits and the smooth curves represent the respective PDFs for the two models.} } 
\label{fig:4}
\end{figure*}

While the two models used here have different distributions in orbital parameters, these are only considered in our calculation of degradation factors. It is clear from Fig.~\ref{fig:physical} that the starkest difference between the models are the neutron star masses and total system lifetimes. {Thus, given that we keep all parameter distributions relating to pulsar evolution fixed, neutron star and black hole mass and system lifetime are the main differentiators between the two models. This implies that the detectability of NSBH systems is strongly dependent on constituent masses and the system lifetime, which is governed by the evolution of orbital parameters, from stellar ignition of the progenitors through to the common-envelope phase and inspiral}.

Similar to \citet{Olejak_2021}, where the authors showed the nonlinear sensitivity of modeling the CE phase on the resulting binary merger rates, we here observe a similar effect. While the properties of the resulting NSBH binary population are somewhat different, we observe clear differences in the subsequent upper limits on population number and merger rate. We note that we do not include the third model considered in \citet{Olejak_2021}, M481.B, which was shown to be consistent with the merger rates observed by LIGO for binary black hole, binary NS, and NSBH systems. However, here we aim to focus on just the impact of CE evolution modeling, whereas M481.B also included a revised treatment of mass transfer.

Within our framework we also extend our analysis to include predictions for detections of pulsars within NSBH binaries from future surveys. We conduct the same statistical calculations and simulations outlined in section \ref{pulsar_evol_detect}, but replace the surveys used there with ones currently underway, or planned in the near future, namely the DSA \citep{2019BAAS...51g.255H}, FAST \citep{2021ApJ...915L..28P}, and MeerKAT \citep{2024MNRAS.531.3579T} surveys. The model specifications for these surveys are also outlined in Table~\ref{tab:1}. Given our current estimates for the rate of merging NSBH systems in the Galaxy, we provide predictions for these surveys, given in Table~\ref{tab:2}. Our preliminary analysis places the upper limit of detections of pulsars in NSBH binaries at 18 for the DSA, 12 for the FAST, and of order one for MEERKAT. Our initial results imply that CE treatment in population synthesis codes can have a non-negligible impact on the probability of observing pulsars in these systems, which is also dependent on the specific pulsar survey. For example, while there is little difference between the two evolutionary models considered here in terms of their observability by the MeerKAT survey, there is some difference between them for the DSA and FAST pulsar surveys. {\citet{debatri_popsynth_fullpsrpop_2021} and \citet{Shao_2018} conducted studies similar to the preliminary one presented here, and also produced estimates for the current number of pulsars in the Galaxy, at 50--2000 and 3--80, respectively. These are far below our estimates here, with upper limits 9,300 and 11,200 for models 380 and 480. Further, \citet{debatri_popsynth_fullpsrpop_2021} predicts that FAST will observe 0--40 pulsars in NSBH binaries, while we find that FAST will detect at most one. \citet{Shao_2018} predicts that FAST will observe at most eight. These differences likely arise from the differences in the employed population synthesis codes and the subsequent radio selection simulations. While many of the pulsar evolution prescriptions are similar between these studies, \citet{Shao_2018} did not include a prescription for reduced detectability causes by Doppler smearing, and \citet{debatri_popsynth_fullpsrpop_2021} used a fitting formula assuming the same BH and NS masses and orbital inclination angle for all binaries. This differs from our treatment, following the techniques described in \citet{Pol_2021}, while we also consider the constituent masses read from \startrack{}. Further, from the top left panel of Fig.~\ref{fig:physical}, very few binaries feature low eccentricities, which will affect their detectability. However, a more systematic comparison of our preliminary results to those in these previoius studies would require direct comparisons with the models of CE evolution between the population synthesis codes, which is outside the scope of this work. Finally, our results suggest} that a more rigorous study may be conducted, where a large number of population synthesis models are considered, varying only how the CE phase is modeled. Our framework could then be used to predict results from future or current observations.

\section*{Acknowledgments}

T.P.J. acknowledges support from NASA FINESST-80NSSC23K1437, and from West Virginia University through the STEM Graduate Fellowship.
N.P. and M.A.M. acknowledge support from the NANOGrav Physics Frontiers Center funded by the National Science Foundation (NSF) under award numbers 1430284 and 2020265. M.A.M. and D.R.L. acknowledge the kindness and many fruitful conversations with K.B. during his highly productive and collaborative career.

\label{lastpage}
\end{document}